
 \input phyzzx

\def\to{\rightarrow}
\tolerance=500000
\overfullrule=0pt

\def\su{s_{1}}
\def\sd{s_{2}}
\def\st{s_{3}}
\def\no{n_{1}}
\def\nd{n_{2}}
\def\nt{n_{3}}
\def\ep{\epsilon_{+}}
\def\em{\epsilon_{-}}
\def\epc{\overline{\epsilon}_{+}}
\def\emc{\overline{\epsilon}_{-}}

\def\rpol{A.M. Polyakov: Mod. Phys. Lett.
{\bf A6}(1991) 635}
\def\rlian {B.H. Lian and G.J.
Zuckermann:
Phys. Lett. {\bf 226}(1991) 21.
 P. Bouwknegt,
J.McCarthy and K. Pilch: CERN-TH.
 6192/91}
\def\rwitten{E. Witten:
 IASSNS-HEP-91/51}
\def\rklebpol{I.R. Klebanov and A.M.
 Polyakov:
Mod. Phys. Lett. {\bf A6} (1991) 3273}
\def\rmmod{G. Moore and N. Seiberg:
Rutgers
and Yale preprint, RU-91-29. YCTP-P19-91.
 J. Avan and  A. Jevicki: Brown
 preprints
BROWN-HET-824 and 839.
 D. Minic, J. Polchinski and Z. Yang:
 Texas preprint UTTG-16-91.
 S.Das, A. Dhar, G. Mandal and S. Wadia:
 IASSNS-HEP-91/52 and
91/72.}

\Pubnum={\vbox{ \hbox{CERN-TH.6379/92}
\hbox{FTUAM-92-02}}}
\pubnum={CERN-TH.6379/92}
\date={January, 1992}
\pubtype={}
\titlepage

\title{PERTURBING THE GROUND RING OF\break
 2-D STRING THEORY}
\vskip 1.0cm
\author{J.L.F. Barb\'on }

\address{Theory Division, CERN\break
 CH-1211 Geneva 23, Switzerland
\break\break and \break\break
Departamento de F\'isica Te\'orica \break
Universidad Aut\'onoma de Madrid \break
Cantoblanco  28049, Madrid \foot{\rm
{Permanent address}}}

\abstract{ We use free field techniques
in D=2 string theory to
calculate the perturbation of the special
 state algebras when
the cosmological constant is turned on.
In particular, we find
that the ``ground cone" preserved by the
 ring structure  is promoted to
a three dimensional hyperboloid as
 conjectured by Witten.
On the other hand, the perturbed (1,1)
 current algebra of moduli
deformations is computed completely,
and no simple geometrical
interpretation is found.
We also quote some facts concerning the
Liouville/ matrix model dictionary in
this class of theories.}

\endpage

\pagenumber=1

\chapter{Introduction}
Many efforts have been focused in
recent times on the study
of c=1 non critical strings, D=2
critical strings in disguise.
They are interesting as toy models
for the study of the classical
open problems in string theory, like
background dynamics, string
field theory and other non perturbative
 methods. The matrix model
formulation is extremely powerful in
actual calculations, and allows
for a complete solution of the model
to all orders in perturbation
theory, and even more than that, it
provides a framework
in which one can study non perturbative
 effects. Despite the
successes of matrix models as  calculational
tools, we still need
the usual continuum formulation in order
to settle the physical
picture, like the spectrum and the role
 of the symmetries.

Naively one could expect decoupling of
two chains of string oscillators,
just as in the critical 26 dimensional
 string, leaving only the
center of mass degree of freedom (the
tachyon, which is massless in D=2).
However, a careful analysis shows the
 existence of an infinite
discrete set of massive oscillator
 degrees of freedom which fail to
decouple. These are the so called
singular states, first identified
in the matrix model \REF\gross{D.J.
 Gross, I.R. Klebanov, and
M.J. Newman: Nucl.Phys. {\bf 350}
(1991) 621}[\gross] and resonant
Liouville
correlators \REF\pol{\rpol}[\pol],
and  subsequently studied in full
 detail
by
the BRST procedure
 \REF\lian{\rlian}[\lian]
. Choosing an appropiate gauge one can
write them as a singular vector
in c=1 Virasoro times the standard
gravitational
dressing of Liouville: $$
W_{s,n}^{\pm} = V_{s,n}
e^{\sqrt{2}(1\mp s)\phi}
\eqn \uno
$$

where $ V_{s,n} $ can be computed
most easily using the SU(2) Kac-Moody
algebra at c=1,
$$
V_{s,n} \sim \left( \oint
e^{-i\sqrt{2}\psi}\right)^{s-n}
e^{i\sqrt{2} s \psi}
\eqn \dos
$$
for $s=0, 1/2, 1, ...; |n|<s $,
the $ SU(2)$ indices,
and $\psi$ denotes the (non-compact)
$c=1$ free boson. For the compact
case we still have
this $SU(2)$ structure at radius
$ R = 1/\sqrt{2}$ and
$s =$ integer.

Remarkably enough, this currents
enlarge the symmetry of the theory
to $ W_{1+\infty} $
\REF\witten{\rwitten}[\witten].
\REF\klebpol{\rklebpol}[\klebpol]
\REF\mmod{\rmmod}[\mmod]
$$
[W_{s,n},W_{s',n'}] = (s n' - s' n)
W_{s+s'-1,n+n'}
\eqn \winf
$$

The role of the singular states is
fully clarified in the beautiful
analysis
by Witten [\witten], who found that
the
exotic (non standard ghost number)
operators in the BRST cohomology
generate a ring structure under OPE.
 The
previously found special currents
 act on this operator ring as area
preserving vector fields, that is,
the ground ring provides a
representation space for $W_{1+\infty} $.
 The overall picture is the
following: for the free case
(no cosmological term in the action) the
so called ground ring of ghost
 number 0, spin (0,0) operators has four
generators, $
a_{i} $ $ i = 1,2,3,4 $, satisfying the
 condition $ a_{1}a_{2} - a_{3}a_{4} =
0$, which defines a 3  dimensional
cone in $ a_{i} $
 space. Out of the $ W $
currents and  the $ a_{i} $ we can
 construct (1,0)
 and (0,1) (non chiral)
currents $ J_{s,n,n'}(z,\overline{z})
$ ($n=n' $ for the non compact
 case) realizing
 an
algebra of area preserving vector
fields on the ground cone.
In fact, $a_{1} $
and $a_{2} $ correspond to  the
coordinates of the fermionic
 phase space in
the matrix model formulation,
while $\partial\psi - \overline
{\partial\psi}$, the operator
measuring the left-right momentum
difference, corresponds to the
time coordinate in the matrix model.
In the uncompactified
case, all operators should be
invariant under its action, and the
previous structure collapses to
the area preserving
diffeomorphisms of the $a_{1},a_{2}$
 plane ($W_{1+\infty} $) that
also preserve the locus $ a_{1}a_{2}
 =0 $, which in turn is related
to the Fermi surface of the matrix
 model.

It was conjectured in [\witten]
 that the non zero cosmological constant
correction to this structure amounts
 simply to change the cone into
a more general quadric, $ a_{1}a_{2}
 - a_{3}a_{4} = \mu $, and the
Fermi surface into $ a_{1} a_{2} = \mu $.
 In this case the currents
$J, \overline{J}$ would act as volume
 preserving vector fields of
this quadric.

On the other hand, (1,1) currents
 constructed as left-right producs
of W's,
$$
Z_{s,n,n'}^{\pm} = W_{s,n}^{\pm}
\overline{W}_{s,n'}^{\pm}
\eqn \zets
$$
($n=n' $  for the non compact model)
 correspond to the infinitesimal
moduli related to marginal deformations
 of the free D=2 string.
Their current algebra is
$ (W_{1+\infty})^{2} $ for the
free case,
and the structure constants determine
 the beta functions to second
order in the couplings. It was conjectured in
 [\klebpol] that for $\mu \neq 0$
this $ W_{1+\infty} $ deforms into
some $SL(2,R)$ tensor operator
algebra ${\cal T} (\lambda) $
($ \lambda = \rm{casimir} $). In particular
$ {\cal T} (\infty) $ is the
 algebra of area preserving diffeomorphisms
 of a two
dimensional hyperboloid \REF\wu{E. Bergshoeff,
M.P Blencowe and K.S. Stelle:
Comm. Math. Phys. {\bf 128} (1990) 213.
C. Pope, L. Romans and X. Shen:
Nucl. Phys. {\bf B339} (1990) 191.}
[\wu].

In the present work, we try to get some
insight
on this conjectures. In particular, we
certainly find a deformation of both the
Fermi surface and the ground cone, while a
geometrical interpretation of the moduli
current algebra is still elusive, in spite
of the fact that the complete structure
constants present a relatively simple
pattern, after cancelation by the
Liouville
sector of most of the tachyon-like terms.
Also,
an
interpretation is given of what
the perturbed algebra
of the $J,\overline{J}$ currents would be.
Unless otherwise stated, we consider the
non-compact model in concrete computations.

\chapter{Perturbation of the ground ring}

Let's first fix our notation. $X =
( \psi,\phi) $ denotes the
matter/Liouville coordinates and
the total e.m. tensor is
$$
T_{tot}(z) = -2b(z)\partial c(z) - c(z)
\partial b(z)
- {1\over 2}\partial X_{\mu} \partial X^{\mu}(z)
+ {i \tilde{Q}^{\mu} \over 2} \partial^{2}
 X_{\mu}
\eqn \tensor
$$
where $\tilde{Q}=(\alpha_{0},-i Q)$ is
the background charge
vector, given by $ (0, -2\sqrt{2} i )
 $ for $c=1$.
We assume the
euclidean metric and consider
 imaginary momenta in the Liouville
direction, $ p = ( k, -i\beta)$.

Physical (0,0) ghost number 0 chiral
 operators have the following
``tachyon + photon" form
$$
x_{\pm}(z) = {\cal O}_{{1\over 2},
\pm {1\over 2}}(z)
= \left( j(z) + i\xi_{\mu}^{\pm}\cdot
\partial X^{\mu}(z)\right)
e^{ip^{\pm}\cdot X(z)}
\eqn \xx
$$
$j(z)=c(z)b(z)$ is the standard ghost
 current and the allowed
momenta/polarizations are given by
$p^{\pm} = (\pm 1/\sqrt{2},
i/\sqrt{2}), \xi^{\pm}
= (\pm 1/\sqrt{2}, -i/\sqrt{2})$
for the special states at $c=1$. We
 can combine
this basic chiral fields in four ways
to form full quantum operators:
$$
a_{+}= a_{1}=x_{+}\overline{x}_{+}
\;;\;
a_{-}= a_{2}=x_{-}\overline{x}_{-}
\;\;\;;\;\;\;
a_{3}=x_{+}\overline{x}_{-} \;\;;\;\;
a_{4}=x_{-}\overline{x}_{+}
\eqn \aes
$$
and the quadratic relation defining
the cone
$a_{1}a_{2}-a_{3}a_{4}=0$
arises automatically because of
 chiral factorization, (in the
non-compact model we must balance
matter momenta between $a_{3}$ and
$a_{4}$). However, in
 a fully interacting Liouville
 sector we do not expect this
operators to have a simple chiral
 square root, allowing for Witten's
hyperbolic deformation
 $a_{1}a_{2}-a_{3}a_{4}=\mu$.
One should understand this
 equation as valid in correlation
 functions
or under the perturbed OPE of
 currents acting on $a_{i} $
 polynomial
functions defined on the quadric.
 Indeed, we'll explicitly calculate the
normalized correlator
$$
{1\over Z(\mu)}\langle(a_{1}a_{2}-a_{3
}a_{4})\rangle_{\mu}
\sim \mu
\eqn \co
$$
with an appropiate renormalization
 of the $a_{i} $. To evaluate \co\
when we add $\mu$ requires a suitable
definition of the product involved
i.e. a regularization.
At this point
we can take advantage of the fact
that the $ a_{i} $ operators enjoy a
ring structure with a singularity
free OPE,
at least up to BRST trivial operators.
Then a natural regularization
would be just point splitting.
$$
\lim_{\epsilon_{\pm}\to 0}\left(a_{1}
(\epsilon_{+})a_{2}(-\epsilon_{-})
-a_{3}(\epsilon_{+})a_{4}
(-\epsilon_{-})\right)=
$$
$$
\lim_{\epsilon_{\pm} \to 0}
 x_{+}(\ep)x_{-}(-\em)
\left( \overline{x}_{+}(\epc)
\overline{x}_{-}
(-\emc)- \overline{x}_{-}(\epc)
\overline{x}_{+}(-\emc)\right)
\eqn \split
$$
Remarkably enough, this limit is
 non zero when we allow the $a_{i} $
to interact with the cosmological
 constant operator.

Before addressing this point, let's
 consider some general properties
of correlation functions of $a_{i}$
 operators. To be specific, we look
at correlators of the form ($Q={2\over
 \gamma}+\gamma$):
$$
{\mu^{s} \Gamma(-s)\over Z(\mu)}
\langle a_{+}(\ep)a_{-}(\em)
\left(\int e^{\gamma \phi}
\right)^{s}
\prod_{j=1}^{N} \int T_{p_{j}}
\rangle_{\mu=0}
\eqn \cos
$$
where we have integrated out
the Liouville
zero mode in the standard
way, and introduced N
tachyons $T_{p_{j}}=e^{ip_{j}
\cdot X},p_{j}=
(k_{j}, -i\beta_{j})$ for
generality. We consider only ``bulk"
correlators, for which
$s= {1\over \gamma}(Q-\beta_{+}-
\beta_{-}-
\sum\beta_{j})$ is a
 positive integer.
In that case we interpret the infinite
in $ \Gamma(-s) $
as a volume divergence
associated with  the Liouville coordinate,
it is natural then to
susbstitute $ \Gamma(-s) \rightarrow
(-)^{s}
|log \mu|/s! $. Taking into account
the scaling of the c=1 partition
function,
$ Z(\mu) \sim \mu^{2}
|log\mu| $
we can write the correlator of
interest in \cos\ as
$$
F(\ep,\em;\mu)= (-)^{s}
{\mu^{s-2}\over s!}
 \int
\prod_{i=1}^{s}d^{2}w_{i}
\prod_{j=1}^{N}d^{2}z_{j}
{\cal A} \;\;\overline{\cal A}
\eqn \aa
$$
the unintegrated chiral
 amplitude is given by
$$
{\cal A}(\ep,-\em,w_{i},z_{j})\equiv
{\cal A}(Z_{\alpha})=
\langle x_{+}(\ep)x_{-}(-\em)
\prod_{\alpha\not= \pm}
e^{iP_{\alpha}\cdot X(Z_{\alpha})}
\rangle
\eqn \a
$$
where $P_{\alpha}=(p_{+},p_{-},
\overbrace{p_{\gamma},...,p_{\gamma}}
^{s},
p_{j})\;\;,\;\;\; p_{\gamma}=(0,-i\gamma)$.

The resulting free $\mu =0$
 correlators are performed using
free field
contractions based on the propagator
$$
\langle X^{\mu}(z) X^{\nu}(w)\rangle
 = - \delta^{\mu\nu}
log(z-w)
\eqn \prop
$$
The ``photon" terms are treated in
the standard way in critical string
amplitudes, and the integrals above
should be understood
 as continued from $ c < 1$.
In this way, using the cosmological
 constant operator
$$
\mu\tilde{T}_{\gamma} =
 \mu{e^{\gamma\phi}\over\Delta(-\rho)}
;\;\;\;
\Delta(x) \equiv {\Gamma(x)\over
\Gamma(1-x)};
\;\;\;\;\rho \equiv -{\gamma^{2}
\over 2}
\eqn \cosmo
$$
in the $c \rightarrow 1$ limit we
effectively insert the true microscopic
operator (in the terminology of
\REF\sei{N. Seiberg: Prog. Theor.
Phys. Suppl.
{\bf 102} (1990) 319.}[\sei] $\mu\phi
e^{\gamma\phi}$).In order to actually
calculate the
integrals we must tune the momenta
$ p_{j}, p_{\pm} $ in such a way
that charge conservation is achieved
$$
p_{+}+p_{-}+s p_{\gamma} +
\sum_{j=1}^{N} p_{j} =
\tilde{Q}
\eqn \carga
$$
with s a positive integer.
For $N=0$, this conditions
plus the $(1,1)-$operator conditions
uniquely
fix the momenta,
for  given
$\alpha_{0}$ and $s$. Thus in general
 there is not enough freedom to tune
the
singular operator's momenta. Although
 this operators
only exist as BRST non
trivial classes for particular momenta,
 they
are generically Virasoro
primaries, in contrast with
the case of (1,1) discrete
currents. This is easily
seen by performing the OPE of $
 x_{\pm}$ with the
total e.m. tensor \tensor\ .
They are (0,0) Virasoro primaries if
$$
L_{1}x_{\pm}=(\xi^{\pm}
\cdot(p^{\pm}-\tilde{Q})-3)
x_{\pm}=0
\;\;\;\;\;\;\;
L_{0}x_{\pm}=({1\over 2}p^{\pm}\cdot
(p^{\pm}-\tilde{Q}) +1)
x_{\pm}=0
\eqn \eles
$$
where the factor of 3 comes from
the ghost anomaly
on the sphere. The
important point is that these are
non-homogeneus
equations in $p$ and $\xi$,
thus having generically non-trivial
 solutions.
This is an ilustration of
the fact that BRST cohomology is rather
subtle
for those operators with ghost content.
In particular, the
$x_{\pm}$'s have an associated
current $U_{\pm}(z)=b_{-1}(x_{\pm})=
b(z){\rm exp}(ip^{\pm}\cdot X(z))$ ,
but by fixing again this current in the
standard way,
we obtain $c(z)U_{\pm}(z)=
j(z){\rm exp}(ip^{\pm}\cdot X(z))$. It
 seems that
we have lost the ``photon" part.
In fact, the 1-1 correspondence between BRST
closed states and (0,0) Virasoro
primaries occurs only for ghost free fixed
currents of the form
$c(z)\times (1,1)\rm{current}(\psi,\phi)
$ as for critical strings.

The net outcome of this disgression is
that we can tune the
polarizations $\xi $ in order to have
Virasoro
primary operators
along our s=integer trajectory $c
 \rightarrow 1^{-}$.

\section{Fermi surface}

The KPZ scaling of the product
$a_{+}a_{-}$ is
$\mu^{3}$. This means that $a_{+}a_{-} \sim
\mu$ in normalized correlators like \cos\
in agreement with the expectations
in [\witten]. Of course we must show
that the
proportionality constant (the integral
in
\aa\ ) is non vanishing.

Let's calculate the chiral
amplitude $\cal A$ in \a\ .
We may split it in two terms
${\cal A}_{0} + {\cal A}_{1}$:
a ``tachyon" part,
$$
{\cal A}_{0}= \langle j(\ep)j(-\em)
\rangle
\langle \prod_{\alpha} e^{iP_{\alpha}
\cdot X(Z_{\alpha})}
\rangle=
$$
$$
={1 \over (\ep + \em)^{2}} \prod_{\alpha
<\beta}Z_{\alpha\beta}
^{P_{\alpha}\cdot P_{\beta}} ;
\;\;\;\;\;(Z_{\alpha\beta}=
Z_{\alpha}-Z_{\beta})
\eqn \acero
$$
and a ``photon" part,
$$
{\cal A}_{1}= \langle i\xi^{+}\cdot
\partial X
e^{ip^{+}\cdot X(\ep)} i\xi^{-}\cdot
 \partial X
e^{ip^{-}\cdot X(-\em)}
\prod_{\alpha \not= \pm}e^{iP_{\alpha}
\cdot X(Z_{\alpha})}
\rangle =
$$
$$
=\{{\xi^{+}\cdot \xi^{-}\over
 (\ep+\em)^{2}}+
(f^{+}\cdot \xi^{+})(f^{-}\cdot
\xi^{-})\}
\prod_{\alpha<\beta}Z_{\alpha\beta}
^{P_{\alpha}\cdot P_{\beta}}
\eqn \auno
$$
where
$$
f^{\pm}={\pm p_{\mp}\over {\ep+\em}}
+\sum_{k=1}^{s}{ p_{\gamma}\over{\pm
 \epsilon_{\pm}-w_{k}}}
+\sum_{j=1}^{N}{ p_{j}\over{\pm
\epsilon_{\pm}-z_{j}}}
\eqn \efes
$$
Taking into account that $\xi^{+}
\cdot\xi^{-}\rightarrow -1 $
and $\xi^{\pm}\cdot p^{\mp}
\rightarrow 0$  as $c\rightarrow
1$, our final result, up to terms
 vanishing in this limit, is
$$
{\cal A}(Z_{\alpha}) = (h^{+}
\cdot\xi^{+})(h^{-}\cdot
\xi^{-})
\prod_{\alpha<\beta}|Z_{\alpha\beta}|^
{ P_{\alpha}
\cdot P_{\beta}}
$$
$$
h^{\pm} =\sum_{\alpha\not= \pm}
{P_{\alpha}\over
\pm \epsilon_{\pm}-Z_{\alpha}}
\eqn \vuelta
$$
After changing variables
$Z_{\alpha}\rightarrow (\ep-Z_
{\alpha})/
(\ep+\em)$ in the
integral \a\ , and  making repeated
use of the on shell conditions,
 (charge
conservation  and  conformal
 weights
 0 and 1 for fixed and moving
operators
resp.),  we find that the dependence
in $\epsilon_{\pm}$
cancels as it should be,
since we are dealing with the two
 point function of (0,0)
operators. In fact, for the particular
 case in which
all $P_{\alpha}$ are equal to each
 other ($P_{\alpha}
= p_{\gamma}$) the problem is simply
reduced
to a tachyon amplitude, because in
this case we
can trade the $h^{\pm}$ terms by a
differential
operator
$$
{\cal A}(Z_{\alpha}) = -(\ep +
\em)^{p^{+}\cdot p^{-}}
{\xi^{+}\cdot p_{\gamma}\over
\xi^{+}\cdot p_{\gamma}}
\partial_{\ep}\partial_{\em}
\prod_{\alpha \not= \pm}(\ep -
Z_{\alpha})^{p^{+}\cdot p_{\gamma}}
(-\em - Z_{\alpha})^{p^{-}\cdot
 p_{\gamma}}
\prod_{\alpha<\beta \not=\pm}
Z_{\alpha\beta}^{P_{\alpha}
\cdot P_{\beta}}
\eqn \taqui
$$
Inserting this expression in \aa\ we
end up
with the by now famous B-9 integral
of \REF\dotfat{V.S. Dotsenko and
V.A. Fateev: Nucl. Phys. {\bf 251}
 (1985) 691}[\dotfat],
$J(\alpha,\beta,\rho)$ . In our case
$\alpha=\beta\rightarrow 1$ and
 $\rho = -\gamma^{2}/2
\rightarrow -1$. Using the
Liouville charge relation we have,
 $\alpha = -1-\rho
(N+s-1)$ and after some algebra we
 get
$$
J_{s+N}(\alpha,\alpha,\rho) \sim
\Delta(0) \prod_{i=1}^{s+N-1}
\Delta(-i\rho)
\eqn \div
$$
We find a divergent result!.
In fact , this divergence
along all the $c\rightarrow 1^{-}$
trajectory is due to the volume of
the dilatation group, which remains
from $SL(2,C)$ after fixing two points.
In order to divide out this volume
we can factor the Haar measure for
dilatations,
$$
\int d^{2}\gamma\int d^{2}\beta
{d^{2}\alpha\delta^{2}(\gamma)
\delta^{2}(\beta-1) \over
|\alpha -\beta |^{2}
|\alpha - \gamma |^{2}
|\beta -\gamma |^{2}}
 =
{d^{2}\alpha \over {|\alpha|^{2}
|1-\alpha|^{2}}}
\eqn \haar
$$
and fix, say $\alpha = \infty $.
 The net result is
just the elimination of one
of the integrals:
$$
{J_{s+N}\over Vol(\rm{dil})} =
J_{s+N-1}(\alpha, \alpha,\rho)\sim
\Delta(-\rho)^{N}
\Delta(1+(s+N)\rho)
\eqn \jotamenos
$$

Upon using once more the Liouville
charge relation
we have $N + s = 3$.
If we also renormalize the new N
tachyons by the
usual leg factor
$\Delta(-\rho)^{-1}$, and the product
 $a_{+}a_{-}$ by
$\Delta(-3\rho)$ we finaly arrive at
$$
{1 \over Z(\mu)} \langle a_{+}a_{-}
 \tilde{T}_{\gamma}
^{N}\rangle_{\mu} \sim {(-)^{N}
\over(3-N)!}\mu^{1-N}
\eqn \fermis
$$
The renormalization of the $a_{\pm}$
operators
is divergent in the $c=1$ limit.
Maybe it could be interpreted as a
kind of
``leg pole" similar to the ones
occurring in the special tachyon
correlators.
In consecuence we see that
the properly renormalized product
$a_{+}a_{-}\equiv
a_{1}a_{2}$ is proportional
to $\mu$ inside correlators
of cosmological constant operators.
Note
however that the proportionality
constant in
\fermis\ depends on the number N
of tachyons.
This could be due to the effect of
 non-decoupled
BRST conmutators. After all, as was
observed in \REF\kut{D. Kutasov, E.
Martinec and N. Seiberg: Princeton and
Rutgers preprints, PUPT-1293, Ru-91-49.
I.R. Klebanov: Princeton preprint
PUPT-1302}
[\kut], the cosmological
 constant
is in the same module (representation)
of the ring as the more complicated
discrete states. In this module the
free relation $a_{+}a_{-}=0$ is not
supposed to be valid. However, since
$$
\langle \tilde{T}_{\gamma}^{N+1}
\rangle_{\mu}
\sim \left({\partial\over \partial
\mu}\right)
^{N-2}\mu^{s+N-2} \sim {\mu^{3-N}\over
(3-N)!}
\eqn \curro
$$
we find the following $operator$
identification
inside correlators of cosmological
constants
$$
a_{1}a_{2} \sim \mu \tilde{T}_{\gamma}
\eqn \guapa
$$

\section{Ground cone}

Now we are ready to check the deformation
of the ground cone. According to
\split\ we
must calculate
$$
C(\mu)=\lim_{\epsilon_{\pm}\to 0}
{1\over Z(\mu)}
\langle\left(a_{1}(\ep)a_{2}(-\em)
 - a_{3}
(\ep)a_{4}(-\em)
\right)\prod_{j}\int T_{p_{j}}
\rangle_{\mu}
\eqn \cmu
$$
In terms of the chiral amplitudes
introduced
before,
$$
C(\mu) \sim \mu^{1-N}\int
\prod_{i=1}^{s}d^{2}w_{i}
\prod_{j=1}^{N}d^{2}z_{j}
{\cal A}(\ep,-\em)\left(
\overline{\cal A}(\epc,-\emc) -
\overline{\cal A}
(-\emc,\epc)\right)
\eqn \outra
$$
and all we need is to compute
the braiding between
the two special
operators in the antiholomorphic
 part of the
unintegrated correlator. In the
general case there
is no simple expression for this
braiding. But
again for the case $P_{\alpha} =
p_{\gamma}$ we
find a big simplification. The
 corresponding
chiral amplitude is given by
\taqui\ , whose
braiding phase is $(-)^{p_{+}
\cdot p_{-}}$,
just $-1$ at $c=1$. In consecuence,
 \cmu\
is exactly twice the integrals in
the previous
section.

Such a result might seem paradoxical
since
we already argued that the final
 amplitudes
are independent of $\epsilon_{\pm}$.
 However,
note that the braiding is taken
 $before$
integrating the moving operators,
this is
technically the explanation of
 why
point-splitting actually works.

To conclude, we have shown that,
 assuming the same
renormalizations as in the previous
section
$$
{1\over Z(\mu)}\langle\left(a_{1}
a_{2}-a_{3}a_{4}\right)
\tilde{T}_{\gamma}^{N}\rangle_{\mu}
\sim \mu^{1-N}
\eqn \ya
$$
So, at least in the presence of
the
cosmological constant operator,
we have the
identification
$$
a_{1}a_{2}-a_{3}a_{4} \sim \mu
 \tilde{T}_{\gamma}
\eqn \chula
$$

\chapter{Perturbed current algebra
 of moduli deformations}

Generally speaking, (1,1) currents
are associated to marginal
 deformations
of the theory, at least on a
first approximation. Trully
marginal deformations, (so called
integrably marginal) correspond
to (1,1)
currents  to all orders in
 perturbation theory. In the
uncompactified
model, the moduli associated
 with infinitesimal deformations
of the 2-D free string are
$$
Z_{s,n}^{\pm} = W_{s,n}^{\pm}
\overline{W}_{s,n}^{\pm}
\eqn \zetas
$$

For a general perturbation of the
 form
$$
S \rightarrow S + \sum_{a}
 \lambda_{a}\int Z_{a}
\eqn \pert
$$
their current algebra structure
 constants $K_{a,b}^{c}$
determine the beta functions up
to second
order in the couplings,
$$
\beta^{c} \sim \sum_{a,b}
 \lambda_{a}\lambda_{b}
K_{a,b}^{c}
\eqn \chorra
$$
In was found in [\witten] and [\klebpol]
 that this structure constants
 correspond
to $W_{1+\infty}^{2}$ at the free point
 $\mu = 0$. In fact, it is very easy
to see that perturbation theory in
$\lambda_{a}$ is exact for any simple
deformation of the free lagrangian
involving only one $ Z_{s,n}^{\pm}$
operator. This is due to the fact
that this $Z$ operators carry definite
2-D momentum, so that the perturbative
 expansion of any correlator is
truncated to a single term in the
 same way as in the lagrangian
formulation of Coulomb gas conformal
 field theories (see [\dotfat]).
For more general linear combinations
 of the Z's this is no longer true,
just like the symmetric perturbation
 of screenings in the Coulomb gas
gives Sine-Gordon theory, which is a
massive theory (of course this
remarks only apply in a perturbative
sense; exact evaluation of the
zero mode $\grave {\rm a}$
la Goulian-Li
drastically changes the situation).
 Looking
at the perturbed two point function
of Z operators, one can easily see
that the only non trivial (single)
perturbation is $ Z_{1,0}^{+} =
\partial\psi\overline{\partial\psi}$
 which
corresponds to the well known radial
moduli already  present in the c=1
 matter theory.
In the general case it is much
more  difficult to decide whether
a particular
perturbation is marginal or
not.  For example, the second order
 beta function in \chorra\
 vanishes for
perturbations in abelian subalgebras of
 $W_{1+\infty}^{2} $. Taking
for simplicity deformations by
``microscopic"
(+) operators,
and considering the non compact
 case,
we have the general second
order marginal deformation
$$
{\cal D}_{s,n}=
\sum_{s^{'}} \lambda_{s^{'}}
Z_{s^{'},n{s^{'}\over s}}^{+}
\eqn \lujo
$$
(see [\witten] for an expression
in
 terms of $a_{i}$ operators).

Our aim in this section is to
 calculate
the deformation of the structure
constants  along the marginal
 line
corresponding to the pure
 cosmological
constant perturbation. This
is equivalent
to the computation of
three point functions of $Z$
 operators
$\langle Z^{+}Z^{+}Z^{-}\rangle $
 from which
we can read
off the $K_{++}^{+}$ and the
$K_{+-}^{+}$
structure constants. It was
already  shown in \REF\miao{M. Li:
 Santa Barbara preprint, UCSBBTH-
91-47.}[\miao] the
vanishing of all
$ \langle Z^{+}Z^{+}Z^{+}\rangle$
correlators. So  we just consider
 the
$ K_{++}^{+}, K_{--}^{+}, K_{--}^{-}$
cases.

The chiral vertex operators are
 constructed
 as in \uno\
where we normalize the matter operators
 by the Condon-Shortley convention,
$$
V_{s,n} = \left[(s+n)!\over
{(s-n)!(2s)!}\right]^{1/2}
\left(\oint
J_{-}\right)^{s-n} V_{s,s}
\eqn \vform
$$
$V_{s,s} $ is just the usual
 vertex operator
${\rm exp}(i\sqrt{2} s \psi)
,\;\;
s\in{\bf Z}/2$ and $ J_{\pm} =
 {\rm exp}
\{\pm i \sqrt{2} \psi\},\;\;
J_{3} = i\sqrt{2} \partial \psi
 $ satisfy the
$ SU(2) $ Kac-Moody
current algebra at level $k=1$.
 In this way
$ V_{s,n}$ is proportional
to a polynomial in derivatives
 of the $\psi$
field, times the
vertex $ {\rm exp}\{ in\sqrt{2}\psi\}
 $ carrying
the momentum. The
total two dimensional momentum of
the
$ W_{s,n}^{\pm} $ fields is
$p = (k,-i\beta)=\sqrt{2}
(n,-i(1 \mp s))$.

We are interested in
correlators of the form
$ K_{++}^{+} \sim
<Z_{s_{1},n_{1}}^{+}Z_{s_{2},
n_{2}}
^{+}Z_{s_{3},-n_{3}}^{-} >_{\mu} $.
 After
$SL(2,C)$ fixing and zero
mode integration we have
$$
K_{s_{1},n_{1};s_{2},n_{2}}^{s_{3},
n_{3}} =
(-)^{s}{\mu^{s} \over s!} |log \mu|
 \langle
Z_{s_{1},n_{1}}^{+}(0) Z_{s_{2},n_{2}}^
{+}(1)
Z_{s_{3},-n_{3}}^{-}
(\infty) \left(\int \tilde T_{\gamma}
\right) ^{s}
\rangle_{\mu=0}
\eqn \cs
$$
for $ \tilde T_{\gamma} $ the
 renormalized
cosmological constant
operator as in \cosmo\ . In the sequel,
we will
drop the Liouville volume term
 $ |log \mu |$
when speaking about the current
 algebra
structure constants. Since $s$ is
always
integer for correlators of singular
operators, it is consistent to
integrate
the Liouville zero mode along the
imaginary
axis. In this way we make contact
with the free ($\mu =0 $) case.

We can split \cs\ into matter and
Liouville parts
$$
K_{s_{1},n_{1};s_{2},n_{2}}^{s_{3}
,n_{3}} =
{\mu^{s} \over s!}(-)^{s} \left(
f_{s_{1}s_{2}}^{s_{3}}
C_{n_{1}n_{2};n_{3}}
^{s_{1}s_{2};s_{3}}\right)^{2}
L_{s_{1}s_{2}}^{s_{3}}
\eqn \mm
$$
where $L$ is the Liouville part,
 $C$
is the
usual Clebsch-Gordan
coefficient, and $ f_{s_{1}s_{2}}^
{s_{3}} $ is
the reduced matrix
element in the matter OPE:
$$
V_{\su,\no}(z) \cdot V_{\sd,\nd}(0)
\sim
\sum_{\st = |\su - \sd|}^{\su + \sd}
z^{\Delta_{3} - \Delta_{1} -
\Delta_{2}}
f_{\su \sd}^{\st} C_{\no,\nd;\nt}^
{\su,\sd;\st}
V_{\st,\nt}(0) + ...
\eqn \ope
$$
The kinematics in \cs\ enforces the
 selection rules:
$\nt = \no+\nd ; s=\su+\sd-\st-1=$
positive integer,
which
combine with the Clebsch-Gordan rule
 to give
$\st=\su+\sd-1,..., |\su-\sd|$.
In particular,
we learn
from this constraints that
 $ K_{--}^{\pm} =0 $ also
for $\mu \not= 0$.
Let's explicitly calculate the
structure constants in \mm.

\section{Matter structure constants}

We want to calculate the function
$f_{\su\sd}^{\st} $. For the
free case we have $s=0,
(\st=\su+\sd-1)$
and the result
quoted in [\klebpol] is
$$
f_{\su\sd}^{\st} = - {(2\su+2\sd-2)!
\over
(2\su-1)!(2\sd-1)!}
\eqn \f
$$
Here we follow a similar argument.
Consider the OPE
$V_{\su,\st-\sd}(z)\cdot
 V_{\sd,\sd}(0) $
and identify the
tachyon on the right hand side.
That is, we select
the term
$$
f_{\su\sd}^{\st}
C_{\sd-\st,\sd;\st}^{\;\su\;,\sd;\st}
V_{\st,\st}(0)
\eqn \opp
$$
which corresponds to the leading one
in the following OPE
$$
\sqrt{(\su+\st-\sd)!\over
{(\su+\sd-\st)!(2\su)!}}
\prod_{j=1}^{s+1}\oint
{du_{j}\over2\pi i}
:e^{-i\sqrt{2}
\psi(z+u_{j})}:
:e^{i\su\sqrt{2}\psi(z)}:
:e^{i\sd\sqrt{2}\psi(0)}:
\eqn \opi
$$
After performing the contractions and
writing explicitly
the Clebsch-Gordan coefficient
(fortunately this can
be done in this case, see \REF\ed{A.R.
 Edmonds: Princeton Univ. Press.,
Princeton N.J. (1957)}[\ed]),
 we find
$$
f_{\su\sd}^{\st} =
\left[{(\su+\sd-\st)!(\sd+\st-\su)!
(\su+\sd+\st-1)!
\over (2\su)!(2\sd)!(\su+\sd-\st)!
(2\su+1)!}\right]
^{1/2} I(\su,\sd,\st)
\eqn \ff
$$
and $I(\su,\sd,\st)$ denotes the
integral
$$
I(\su,\sd,\st) = \prod_{j=1}^{s+1}
\oint{dx_{j}\over 2\pi i}
x_{j}^{-2\su}(1+x_{j})^{-2\sd}
\prod_{j<k}^{s+1} (x_{j}-x_{k})^{2}
\eqn \inte
$$
This integral is explicitly
computed in the
Appendix.

\section{Liouville structure
 constants}

The functions  $L_{\su,\sd}^{\st}$in
\mm\
have already
been computed in [\miao].
Here we will just comment on
some delicate points.

The integral representation
for the L's, prior to
$SL(2,C)$ fixing is given by:
$$
{1\over {\rm Vol}(SL(2,C))}
\int
\prod_{k=1}^{3}d^{2}z_{k}
\prod_{i=1}^{s}d^{2}w_{i}
\prod_{k<l}^{3}|z_{kl}|^{2\delta_{kl}}
\prod_{k,i}|w_{i}-z_{k}|^{-2\gamma
\beta_{k}}
\prod_{i<j}^{s} |w_{ij}|^{4\rho}
 \eqn \integrona
$$
Naive substitution of the $c=1$
exponents gives
a divergent integral and the
 appropiate
regularization
is not known from the Liouville
 point
of view. The
reason is that, unlike the case of
the
$a_{i}$ operators
in the previous section, there are
no
(1,1)-discrete
currents in $c<1$. So, the best one
 can do
at this point is to
continue the exponents in
\integrona\ ,
while maintaining
at least $SL(2,C)$ invariance,
leaving the
matter structure constants
 untouched.

It is very easy to show that
 \integrona\ is
independent
of the fixing of $z_{k}$ if
 we have
Liouville-charge
matching:
$$
\sum_{k}\beta_{k} + \gamma s = Q
\eqn \charge
$$
and the following three equations
 are satisfied:
$$
\sum_{l\not= k}\delta_{lk} = \gamma s
\beta_{k} - 2
\eqn \chachi
$$
So, for fixed s and Q, all
 $\delta_{kl}$
are determined by two of the
 $\beta_{k}$,
say,
$\beta_{0}$ and $\beta_{1}$. And we
are left
with the B-9 integral $J_{s}
(\alpha,\beta,\rho)$,
where
$\alpha = - \gamma\beta_{0},\;\;
\beta = -\gamma\beta_{1},\;\;
\rho = - \gamma^{2}/2 $.

The most general ``$c<1$"
 (projective invariant)
regularized trajectory is thus
given by
$$
\rho = \epsilon - 1\;; \;\;\alpha =
a\epsilon + 2\su - 2\;;\;\;
\beta = b\epsilon + 2\sd - 2 \;\;;\;\;
\epsilon
\rightarrow 0
\eqn \trayec
$$
and $ a,b$ are $free$ parameters.
The choice
made
in [\miao] corresponds to the
 particular
case
$a = {4\over3}-\su$ and $b =
 {4\over3}
 - \sd$.
Note that we are changing the
 conformal
properties of the matter part
for $all$
of this trajectories. To see
this, recall
that the $\delta_{kl}$ coefficients
 have the
form
$$
\delta_{kl} = - \Delta_{kl}-\beta_{k}
\beta_{l}
\;;\;\; \Delta_{ij}\equiv
s_{i}^{2}+s_{j}^{2}
-s_{k}^{2}\;,\;k\not= i,j
\eqn \delturra
$$
If we ask the deformation in
 \trayec\ to
keep  the matter weights
 $\Delta_{kl}$
unchanged, then equations
\charge\ , \chachi\
and \delturra\ imply the trivial
 result:
$s_{i}=0$.

Carrying on the standard
 manipulations and
taking the limit in \trayec\
 one finds
a remarkable result: the
regularization dependent
terms can be packed in
$$
S(a,b)=\prod_{i=0}^{s-1}
{(i+b)(i+a)\over(a+b+s-1+i)}=
{\Gamma(a+s)\Gamma(b+s)
\Gamma(a+b+s-1)\over
\Gamma(a)\Gamma(b)
\Gamma(a+b+2s-1)}
\eqn \amb
$$
while the ``bulk" of the
 B-9 formula just
gives the inverse of the squared
integral $I(\su,
\sd,\st)^{2}$ we found in the
 matter part \inte\ .
This is an interesting
phenomenom, reminiscent
of the similar one
 occurring for tachyon
correlators. From this
 point of view, we find
the cancelation of
($f_{\su,\sd}^{\st})^{2}$
 quite natural,
since they
are the $SU(2)$ reduced matrix
 elements
in \ope\
, thus depending on the special
tachyons (i.e.
the highest weights).

\section{Complete structure
 constants}

If we collect the results in
the previous
sections and absorb some trivial
 factors in
the normalization of the operators,
 we arrive
at the following suggestive result
 for the
complete structure constants
$$
K_{\su,\no;\sd,\nd}^{\st,\nt} =
\mu^{s}{(2\st)(2\st + 1)
(\su+\sd+\st+1)^{2}\over
(\su +\sd -\st-1)!} S(a,b)
\Delta(\su,\sd,\st)^{2}
\left( C_{\no,\nd;\nt}^{\su,\sd;\st}
\right)^{2}
\eqn \constantes
$$
where $\Delta(\su,\sd,\st)$ is Wigner's
triangular
function:
$$
\Delta(a,b,c) = \left[{(a+b-c)!
(a-b+c)!
(b+c-a)!\over (a+b+c+1)!}
\right]^{1/2}
\eqn \wig
$$
We still don't know what would be
the geometrical
interpretation of this structure
 constants
(if any). The presence of the
ambiguous
term $ S(a,b)$ complicates very much
the
analysis. In particular, they
 strongly
resemble the structure constants of
${\cal T}(\infty)^{2}$ (see [\wu]),
although no trace of the parity
 selection
rule, $\su +\sd +\st = {\rm odd}$ is
seen.
Besides, an interesting question
about
\constantes\ is its matrix model
interpretation. As pointed out in
\REF\dan{U.H. Danielsson: Princeton
preprint, PUPT- 1301}[\dan],
in spite of the fact that
left-right
matching of momenta ``explains" why
only one $W_{\infty}$ is seen,
 at the
level of correlation functions we
still
find a doubling of the group
factors,
 since
left and right parts of the Z
 operators
contract independently. This puzzle
is
easily solved by realizing that,
according to [\witten], the correct
dictionary between the matrix model
and
the conformal theory is given by
$$
(p+\lambda)^{s+n}(p-\lambda)^{s-n}
e^{2mt}
\leftrightarrow W_{s,n}^{+}(z)
\overline{{\cal O}}
_{s-1,n}(\overline{z}) \equiv
J_{s,n}(z,\overline{z})
\eqn \mmops
$$
for $p$ the momentum conjugate to
the eigenvalue variable $\lambda$.
Since the ${\cal O}'s$ are in the
ring, correlators of $J's$ do not
double the group factors.
However, this raises the question of
how to write the Z moduli in terms
of matrix model variables.

Incidentally, we must say that
 a definite
answer for \constantes\ can be
obtained
from the operator solution  in the
Gervais-Neveu
interacting Liouville theory
\REF\gerv{J.L. Gervais: Int J. Mod.
Phys.{\bf A6} No.16 (1991) 2805}
[\gerv] .In particular,
one can construct the dressing
operators
explicitly in terms of chiral vertex
operators lying in $SL(2,R)$
 representations.
This vertex operators enjoy
the following
remarkable OPE,
$$
e^{-\gamma\su\phi(z,\overline{z})}
\cdot e^{-\gamma\sd\phi(0)}
= \sum_{\st = |\sd -\su|}^{\su +\sd}
|z|^{\Delta_{3}-\Delta_{2}-\Delta_{1}}
e^{-\gamma\st\phi(0)}
+ ...
\eqn \gervope
$$
It is just the free one, but opened to
all $SU(2)$ channels. So in this
formulation
the final result for the moduli
 OPE is given
by
$$
K_{\su,\no;\sd,\nd}^{\st,\nt} = \mu^{s}
\left(
f_{s_{1}s_{2}}^{s_{3}}
C_{n_{1}n_{2};n_{3}}
^{s_{1}s_{2};s_{3}}\right)^{2} \;\;;\;\;
\st= |\su -\sd|+1,...,\su + \sd -1
\eqn \gervc
$$
the only difference with the pure matter
structure constants is the selection rule
at the bottom. In would be
interesting to
understand why \constantes\ and
\gervc\
look so different.

\chapter{Perturbation of
 $W_{1+\infty}$}

The study of the perturbed algebra
 of
volume preserving diffeomorphisms
 along
the same lines as in the previous
 section
faces a fundamental problem. Recall the
expression for the (1,0) and (0,1)
currents,
$$
J_{s,n,n'}(z,\overline{z})= W_{s,n}^{+}
(z) \overline{{\cal O}}_{s-1,n'}
(\overline
{z})
\eqn \j
$$
and similarly for $\overline{J}$.
(${\cal O}_{s-1,n}$ is BRST equivalent
to $x_{+}^{s+n-1}x_{-}^{s-n-1}$).
Their free OPE is just $W_{1+\infty}
\oplus W_{1+\infty}$. However, we
cannot conjugate $J$ and
$\overline{J}$
in the Liouville momentum in order
to get orthogonal two point functions.
The reason being that the operators
analogous to \j\ , but constructed
out of the $W^{-}$ currents, are
unphysical. This is a striking
feature of the free theory, which
already distinguishes between ($+$)
and ($-$) operators. In consecuence,
we cannot establish a simple
correspondence between OPE and
3-point functions.

Indeed, from the matrix model point
of view, nothing
special happens when the cosmological
constant is turned on (in fact, what
is problematic in matrix models is
the limit $\mu \rightarrow 0)$). The
matrix model operators in \mmops\
still satisty $W_{1+\infty}$ at $\mu
\not= 0$.On the other hand,
on general grounds,
for non zero cosmological constant
we don't expect the quantum
operators
to  factorize in in chiral
left/right
terms in a simple way like in \j\ .
As we'll see in
a moment, this observation is
important.

Let's consider the generalized quantum
 currents
\j\ for $\mu\not= 0$. According to the
ansatz in [\witten] they are
associated to a
certain volume preserving vector
 field on
the quadric
$a_{1}a_{2}-a_{3}a_{4}=\mu$
$$
J_{s,n,n'}(\mu)\leftrightarrow
\mu^{s}w_{A_{1},...,A_{2s}
A_{2s+1}A_{2s+2};
A_{1}',...,A_{2s}'}x^{A_{1}A_{1}'}...
x^{A_{2s}A_{2s}'}x^{A_{2s+1}B'}
\epsilon^{A_{2s+2}B}{\partial
\over \partial x^{BB'}}
\eqn \maestro
$$
for AA' the $SU(2)$ indices for the
four coordinates of the quadric. In
the non compact case, we restrict
\maestro\ to those vector fields
invariant under the operator
computing the left-right momentum
difference: $\partial\psi -
\overline{
\partial\psi}\leftrightarrow a_{3}
{\partial\over \partial a_{3}} -
a_{4}{\partial
\over \partial a_{4}}$. Expanding
the conmutator of two of these
vector fields in the
$SU(2)\times SU(2)$
Clebsch-Gordan series we arrive at
the following fusion rules
$$
[J_{\su}]\otimes [J_{\sd}] \sim
\bigoplus_{\st = |\sd -\su |+1}^
{\su +\sd -1}
\mu^{\su +\sd - \st -1}
[J_{\st}] \;\;;
\;\;\;\su +\sd +\st = {\rm odd}
\eqn \fusrul
$$
(the odd selection rule is due to
 parity
conservation). This algebra contracts
 to
$W_{1+\infty}$ as $\mu\rightarrow 0$.
The
puzzling question is that we $never$
 see
the channels $\st < \su + \sd -1$ in
 the
matrix model. The answer becomes
obvious as we
descend from the homogeneus
variables in
\maestro\ to the minimal set of
 variables
$a_{1}, a_{2},\psi$ for $a_{3}=
\rho e^{\psi}$
$a_{4}=\rho e^{-\psi}$. The set
 of diagonal
vector fields in \maestro\ is
given by
abelian gauge transformations in
the $\psi$
direction, plus those area preserving
vector fields
of the $a_{1}-a_{2}$ plane that
``lift"
correctly to the homogeneus form
in
\maestro\ . That is, those
generated by
hamiltonians of the form:
$$
h(a_{1},a_{2})=(a_{1}a_{2}-\mu)^{m}
h_{0}(a_{1},a_{2})
\eqn \hamil
$$
for $h_{0}$ an arbitrary polynomial
and
$m>1$. After expansion in powers of
$\mu$
we end up with a linear combination
of
hamiltonians in the plane, which we
can
associate with the $free$ currents
\j\ .
So, from this point of view, the
perturbed algebra of volume
preserving
diffeomorphisms arises upon
 perturbation
of the free operators. In the
matrix model
we should take suitable
$\mu$-dependent
linear combinations of the
operators in
\mmops\ in order to activate
 all the
channels in \fusrul\ . All
this structure
is strongly reminiscent of
the similar
phenomenon in interacting
Liouville
theory, where a vertex operator
 is
expanded in a set of chiral
 components,
$$
e^{-\gamma j \phi(z,
\overline{z})} =
\sum_{m=-j}^{j}\xi_{m}^{j}
(z)(-)^{j-m}
\overline{\xi}_{m}^{j}(\overline{z})
\eqn \raiz
$$
and the chiral ``square roots"
 $\xi_{m}^{j}$
transform in an $SL(2,R)$
representation of spin $j$.

\chapter{Conclusions}

We have discussed several  recent
conjectures about the effect of
the Liouville interaction on the
$c=1$ model coupled to two
dimensional
quantum gravity.
In particular, we find that Witten's
 ground
cone is certainly deformed into a
3-dimensional hyperboloid, at
least in
correlation functions. Similar
evidence
for the conjectured deformation
 of the
Fermi surface is also found.
It seems
that the essential ingredient making
possible the alluded deformations is
the lack of chiral factorization
in the
Liouville sector when the
cosmological
constant is turned on. This is
implied
in the calculations on the cone by
the point-spliting regularization
used, and we
propose that the conformal operators
implementing the volume preserving
diffeomorphisms of the quadric
should
be constructed as perturbations
 of the
free fields appearing in the
free case. It would be most
interesting to
make contact
with exact treatments of the
 Liouville
field which already incorporate
this fact,
like [\gerv] \REF\th{E.Braaten, T.
Curtright and C. Thorn: Ann Phys.
{\bf 197} (1983) 365}[\th]. One
would like to see the algebraic
structure encoded in the
$\mu\not=0$ hyperboloid as
a sort of Backlund transform
of the free cone.

Regarding the moduli current algebra,
we explicitly
calculate
the structure constants for non-zero
cosmological constant, although
serious
regularization ambiguities are found
in the Liouville sector.  In spite of
their simple form, we don't find any
geometrical interpretation, which
would be important in order to
 conjecture
string field actions. As a matter of
fact, certain combinations of Z
operators
do satisfy a volume preserving
algebra
under $free$ OPE. This is because Z
 operators
correspond to the hamiltonians
generating
volume preserving vector fields
(J operators).
Free field OPE of the Z's
corresponds to
Poisson brackets squared of
the hamiltonians,
which in turn are equivalent
 to the conmutator
of vector fields. So, if the
picture advocated
in the previous section is
 correct, there
are certain $\mu$-dependent
 deformations
which under free OPE, enjoy
the J's algebra
squared.

{\bf. Acknowledgements}. This work
 arose from
conversations with L. Alvarez-Gaum\'e,
 to whom
I am indebted for his patient tutoring
 and
constant encouragement. I also thank
 C\'esar
G\'omez, Germ\'an Sierra and Javier
 Mas for
stimulating discussions, and the
Cern-TH
division for hospitality. This
 work has been
supported by an FPI-MEC grant.

\chapter{Appendix 1}

In this appendix we give an
explicit formula
for \inte\ :
$$
I(\su,\sd,\st) = \prod_{j=1}^{s+1}
\oint{dx_{j}\over 2\pi i}
x_{j}^{-2\su}(1+x_{j})^{-2\sd}
\prod_{j<k}^{s+1} (x_{j}-x_{k})^{2}
\eqn \intea
$$
the contour integrals are nested
 around 0
in the usual positive sense
(counter-clockwise). Of course,
direct
evaluation of the poles seems
hopeless,
because of the presence of the
 Vandermonde
determinant squared in the
integrand,
which complicates the power
expansion. We
will use analitic continuation in
the
exponents. Consider the integral
$$
I_{\epsilon}(\alpha,\beta;s) =
 \prod_{j=1}^{s+1}
\int_{\gamma_{j}(\epsilon)}
{dx_{j}\over 2\pi i}
x_{j}^{\alpha}(x_{j}-1)^{\beta}
\prod_{j<k}^{s+1} (x_{j}-x_{k})^{2}
\eqn \ineps
$$
The contours $\gamma_{j}
(\epsilon)$ are
homotopic to the ones in \intea and
defined as
$D_{j}(\epsilon)\circ C_{j}
(\epsilon)$,
where
$D_{j}(\epsilon)$ denotes a
small clockwise
circle of radius $\epsilon$
 around 0, and $
C_{j}(\epsilon)= [\epsilon +
 i 0\rightarrow
 1-\epsilon
+ i 0]\circ [1-\epsilon -
i0\rightarrow
\epsilon -i0]$.

For $\alpha=-2\sd$ and $\beta=-2\su$
 we recover
\intea\ for any value of $\epsilon$.
This is because it converges at
$x_{j}=\infty$, so that a change in
 the
orientation of the contour can be
 traded
by the shift $x_{j}\rightarrow
x_{j}-1$.
Now we evaluate \ineps\ in the
 region of
the $\alpha,\beta$ plane were
it is
convergent at 0 and 1. This
 corresponds
to $\alpha,\beta > -1$. So in
this region
we can evaluate the $\epsilon
\rightarrow 0$
 limit by simply neglecting the
small
$\epsilon$-circles $D_{j}
(\epsilon)$, i.e:
$$
I(\alpha,\beta;s)\equiv
\lim_{\epsilon \rightarrow 0}
I_{\epsilon}(\alpha,\beta;s) =
 \prod_{j=1}^{s+1}
\int_{C_{j}}{dx_{j}\over
2\pi i} x_{j}^{\alpha}
(x_{j}-1)^{\beta}
\prod_{j<k}^{s+1} (x_{j}-x_{k})^{2}
\eqn \agggg
$$
where the contours $C_{j}$ are
just the
$\epsilon \rightarrow 0$ limit of
$C_{j}(\epsilon)$.
Now by simple evaluation of
the monodromy:
$$
I(\alpha,\beta;s)= \pi^{-s-1}
 e^{i\pi s\alpha}
(-)^{s+1}({\rm sin}\pi\alpha)^{s+1}
\prod_{j=1}^{s+1}
\int_{0}^{1} dx_{j} x_{j}^{\alpha}
(x_{j}-1)^{\beta}
\prod_{j<k}^{s+1} (x_{j}-x_{k})^{2}
\eqn \mais
$$
The integral in the right hand
 side is
the so
called Selberg-Fattev-Dotsenko
integral
[\dotfat]. After
evaluating it and using $\Gamma(x)
\Gamma
(1-x) = {\pi \over {\rm sin}\pi x}$
we arrive at
$$
(-)^{(s+1)(s+2)\over 2} \pi^{s+1}
({\rm sin}\pi\alpha)^{-s-1}
\left({{\rm sin}\pi(\alpha +\beta)
\over
{\rm sin}\pi\beta}\right)^{s+1}
\prod_{j=1}^{s+1} j!
\prod_{j=0}^{s} {\Gamma(-\alpha
 -\beta
-s-1-j)\over \Gamma(-\alpha - j)
\Gamma(-\beta - j)}
\eqn \chorizin
$$
which can be combined with
\mais\ and
analitically continued back
 to $\alpha
= -2\sd$ and $\beta = -2\su$.
 Finally:
$$
I(\su,\sd,\st) \equiv
I(\su,\sd;s) =
(-)^{s(s+1)\over 2}\prod_{j=1}^
{s+1} j!
\prod_{j=0}^{s} {(2\su +2\sd
-s-2-j)!\over (2\su -1- j)!
(2\sd - 1-j)!}
\eqn \chorizonte
$$

\endpage\refout
\end